\documentclass[runningheads]{llncs}

\usepackage[T1]{fontenc}

\usepackage{newtxtext}
\usepackage{graphicx}
\usepackage{latexsym}

\usepackage{url}

\usepackage{subcaption}
\usepackage{enumitem}

\usepackage{color}

\usepackage{booktabs}
\usepackage{tabularx}
\usepackage{makecell}

\usepackage{float}

\usepackage{ascmac}
\usepackage{tcolorbox}
\tcbuselibrary{breakable}

\usepackage{listings, xcolor}

\let\Return\relax
\usepackage{algorithm}
\usepackage{algpseudocode}
\algrenewcommand\alglinenumber[1]{\scriptsize #1:}
\usepackage{xcolor}

\usepackage{fmtcount}

\usepackage[normalem]{ulem}

\usepackage[hidelinks]{hyperref}
\usepackage{multirow}

\usepackage{bbding}

\usepackage{cite}

\usepackage{setspace}
\newtcolorbox{breakableitembox}[2][c]{ 
  breakable,
  colback=white,
  arc=2mm, 
  title=#2,
  fontupper=\setstretch{0.8}\selectfont,
  parskip=0mm,
}

\makeatletter
	\AtBeginDocument{
	
    }
\makeatother

\newcommand{\countimplications}{
    \def \countimplications{1}
}
\newcounter{implication}
\newenvironment{implication}[1]
{
    \refstepcounter{implication}
    \noindent \textbf{Implication~\theimplication.  \textit{#1}}
}

\countimplications

\newcommand{\conclusionbox}[1]{%
 \vspace{1em}
 \noindent
       \framebox[\textwidth][c]{%
               \parbox[b]{0.9\textwidth}{%
                       { #1}
               }
       }
 \vspace{1em}
}

\newcounter{step}

\newcommand{\RqOne}{
\noindent\textbf{RQ1:} \emph{Does providing flowcharts as visual prompts improve the performance of LLM-based code generation?}
}

\newcommand{\RqTwo}{
\noindent\textbf{RQ2:} \emph{What level of flowchart abstraction is sufficient for effective code generation?}
}

\newcommand{\RqThree}{
\noindent\textbf{RQ3:} \emph{How effective is Few-Shot Learning in code generation using flowcharts?}
}

\newcommand{\RqFour}{
\noindent\textbf{RQ4:} \emph{What is the computational cost of using flowcharts as inputs for LLM code generation?}
}

\newcommand{\AmSLaTeX}{%
 $\mathcal A$\lower.4ex\hbox{$\!\mathcal M\!$}$\mathcal S$-\LaTeX}
\def\BibTeX{{\rmfamily B\kern-.05em{\scshape i\kern-.025em b}\kern-.08em
 T\kern-.1667em\lower.7ex\hbox{E}\kern-.125em X}}

\begin{document}

\title{\textbf{Exploring the Potential of Program Flowcharts on Code Generation Using Multimodal LLMs}\\}

\author{Yuki Toi\inst{1}\orcidID{0009-0000-4567-1607} \and
Tao Xiao\inst{1} \Envelope\orcidID{0000-0003-4070-585X} \and \\
Kazushi Tomoto\inst{1}\orcidID{0009-0009-7455-6304} \and
Masanari Kondo\inst{1}\orcidID{0000-0002-6317-7001}, \\
and Yasutaka Kamei\inst{1}\orcidID{0000-0002-7058-1045}}
\authorrunning{Toi et al.}
%
\institute{Kyushu University, Fukuoka, Japan,
\email{\{toi,tomoto\}@posl.ait.kyushu-u.ac.jp, \{xiao,kondo,kamei\}@ait.kyushu-u.ac.jp}\\
}

\maketitle
\begin{abstract}
In recent years, Large Language Models (LLMs) have made significant strides, leading to the emergence of multimodal LLMs capable of processing diverse inputs such as images and audio. Previous research indicates that the supply of multimodal LLMs with combined textual and visual information improves the automatic code generation capabilities. In software development, diagrams such as flowcharts are widely employed to facilitate tasks like code comprehension. While existing studies investigated the impact of visual inputs on LLMs and the usage of software diagrams, the potential influence of providing flowcharts on multimodal LLM performance remains underexplored. In this study, we generated flowcharts from example solution code for AtCoder problems and provided these visual aids alongside problem statements to GPT-4o for code generation. Our findings demonstrate that integrating flowcharts with problem statements yields performance improvements of up to 10\%. Furthermore, when employing abstracted flowcharts, we observed a trend indicating that increasing levels of flowchart detail correlate with enhanced performance. Additionally, we compared the effectiveness of flowchart provision to Few-Shot Learning approaches. The findings suggest that one-shot learning provides sustainable improvements, whereas two-shot learning results in only minor improvements. Our work highlights the importance of software diagrams in supporting multimodal LLM-driven code generation.

\end{abstract}

\section{Introduction}

Code generation, which involves the automated production of source code from natural language descriptions, represents a cornerstone in the field of automated software engineering. The early approaches to this task were based on heuristic rules~\cite{joshi2003MTT,cohn2010MLR,xiong2017ICSE} and expert systems~\cite{de2008TACAS,gulwani2010PPDP,jha2010ICSE}. However, these foundational techniques were typically rigid and difficult to scale. The advent of Transformer-based Large Language Models (LLMs)~\cite{vaswani2017NIPS} has shifted the paradigm of code generation from the reliance on heuristic rule-based systems to more powerful, scalable approaches that learn patterns from vast code repositories~\cite{jiang2024TOSEM}. This domain garnered substantial interest from academic and industrial experts, as evidenced by the development and widespread adoption of tools such as GitHub Copilot\footnote{\url{https://github.com/features/copilot}}~\cite{yetistiren2022PROMISE,mastropaolo2023ICSE} and ChatGPT\footnote{\url{https://chatgpt.com/}}~\cite{dong2024TOSEM,liu2024TOSEM}.

The practice of software development involves not only textual artifacts, such as source code and documentation, but also essential visual elements such as diagrams and user interfaces.
Therefore, LLMs have been developing, with multimodal LLMs capable of recognizing images and speech in addition to text. GPT-4o,\footnote{\url{https://openai.com/index/hello-gpt-4o/}} developed by OpenAI, is one such multimodal LLM; it can be used interactively with ChatGPT or integrated into a system using the API. These multimodal LLMs have been used for web interface design~\cite{si2025NAACL}, diagramming, and visual testing~\cite{yang2025ICLR}.

The performance of LLMs on code generation tasks can be enhanced through the integration of multimodal support. To this end, Li~et~al.~\cite{Kaixin2024EMNLP} developed the MMCode benchmark to assess the algorithmic problem-solving capabilities of models in visually rich contexts, using GPT-4V\footnote{\url{https://openai.com/index/gpt-4v-system-card/}} and Gemini Pro Vision.\footnote{\url{https://cloud.google.com/blog/products/data-analytics/how-to-use-gemini-pro-vision-in-bigquery?hl=en}}
They observed improved results when a visual context of the problem was supplied alongside its textual description. Nevertheless, the study concludes that successful problem-solving hinges not just on the presence of visual data, but on the advanced capability to deeply comprehend it.

Although extensive research has been conducted on code generation using LLMs, the potential impact of incorporating design diagrams, such as flowcharts, to supplement textual context remains underexplored. The researchers highlight that flowcharts can help developers to understand and design code~\cite{cherubini2007CHI,yatani2009CHI}. We hypothesize that augmenting textual prompts with flowcharts, which visually represent program logic and facilitate comprehension for LLMs, can mimic developers' practices in writing code and will eventually improve the accuracy of code generated by multimodal LLMs.

Therefore, we investigate the effect of incorporating flowcharts on the code generation performance of multimodal LLMs. To facilitate this, we first reused a dataset derived from 756 competitive programming problems on the AtCoder platform, which included 756 test cases and 291 example solutions. From these solutions, we generated the corresponding flowcharts to represent the design or structure of the programming problems. We compared the performance of one of the multimodal LLMs, GPT-4o, on prompts that contain only the textual problem description versus prompts augmented with flowcharts. Then, we explored the impact of visual abstraction by systematically generating flowcharts with varying levels of detail, created by abstracting code blocks based on their indentation depth. Finally, we evaluated and compared the performance and cost when provided with these multi-granularity flowcharts.

Our study has yielded several significant findings related to the incorporation of flowcharts in code generation. First, we observed that providing flowcharts improves code generation accuracy by 4--10\%, with this effect increasing in correspondence with problem difficulty. Second, we determined that abstracted flowcharts using only the most outlier control structures are more effective than the problem description alone. Furthermore, Few-Shot Learning-based prompting strategies for flowcharts result in performance gains for one-shot learning but only marginal gains for two-shot learning. Lastly, the use of flowcharts is cost-effective, achieving a 10\% improvement in accuracy for an additional \$0.000025 per generated instance.
\begin{figure*}[t]
	\centering
	\includegraphics[width=0.86\textwidth]{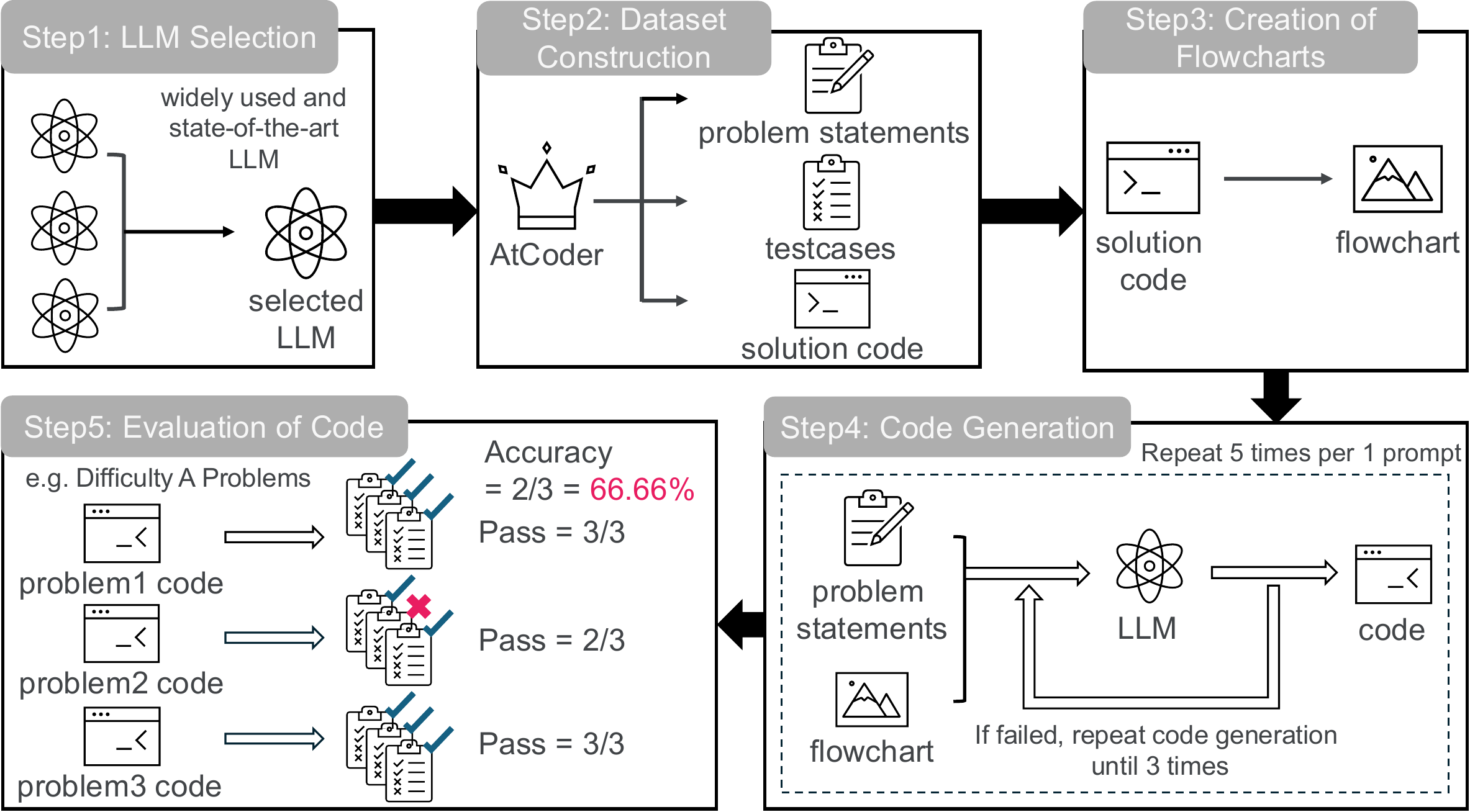}
	\caption{Overview of the experimental settings in our study}
	\label{method_flow}
\end{figure*}

\section{Related Work}
\label{sec:related_work}
This section outlines the related works of this study on the use of LLMs in software engineering, the multimodal capabilities of LLMs, the use of diagrams in software development, and techniques for generating code from flowcharts. Finally, we present the research questions of this study.

\noindent
 \textbf{LLMs in Software Engineering.}
In recent years, LLMs have been used in various tasks in software engineering~\cite{junjie2024TSE,hou2024TSEM,yao2022ACL,fan2023ICSE,zibin2025ESE,celal2025FSE}.
Fan~et~al.~\cite{fan2023ICSE} surveyed the application of LLMs in software engineering.
The results showed that automated code generation using LLMs can improve the productivity of engineers.
Specifically, they found that the group using GitHub Copilot could complete tasks 56\% faster than the group that did not use.
Hou~et~al.~\cite{hou2024TSEM} conducted a systematic review of the literature on LLMs for Software Engineering (LLM4SE). The review showed that LLMs were successfully applied in tasks such as code generation, program repair, and code summarization.
Zheng~et~al.~\cite{zibin2025ESE} surveyed 123 academic papers on LLM4SE and evaluated how LLMs are used and effective in software engineering tasks such as code generation and summarization. The evaluation results showed that LLMs perform well in tasks that require syntactic understanding, such as code summarization and correction.

\noindent
 \textbf{Multimodal Support for LLMs.}
The performance of LLMs has been shown to be improved through multimodal capabilities \cite{Kaixin2024EMNLP,yeasmin2025ICPC}.
Li~et~al.~\cite{Kaixin2024EMNLP} collected data from 10 competitive programming contests, including AtCoder\footnote{\url{https://atcoder.jp/}} and LeetCode,\footnote{\url{https://leetcode.com/}} and proposed and developed a benchmark MMCode to evaluate the code generation capability of multimodal LLMs.
They evaluated several LLMs, including GPT-4V, Gemini and open-source LLMs, and found that the correct response rates of GPT-4V and Gemini Pro Vision were as low as 19.4\% and 5.0\%, respectively.
In particular, the response rates for all models other than the GPT family were less than 5\%.
When comparing the response rates of GPT-4V with multimodal input and GPT-4\footnote{\url{https://openai.com/index/gpt-4/}} with text-only input, they found that GPT-4V performed better on image-related questions for simple tasks such as linear data structures.
However, the performance of GPT-4V declined on questions involving images for difficult tasks such as graph problems.
Yeasmin~et~al.~\cite{yeasmin2025ICPC} analyzed and extracted multimedia information such as images attached to bug reports, and used this information to perform bug identification queries. The results showed that the extended query, which integrates information extracted from images in addition to the traditional text-based query (title + description), improves bug location by 34.06\%. In particular, it was revealed that programming elements in images (code fragments, command outputs, etc.) significantly contribute to search performance.

\noindent
 \textbf{Usage of Flowcharts and UML Diagrams in Development}
\label{subsec:flowchart_usage}
Studies have been conducted on how flowcharts and UML diagrams are used in development \cite{cherubini2007CHI,yatani2009CHI}.
Mauro~et\\~al.~\cite{cherubini2007CHI} interviewed Microsoft developers to investigate how diagrams are used in software development.
The interviews revealed that developers use diagrams mainly in three phases: understanding code, designing code, and communicating during development.
Especially in the code design phase, diagrams drawn on paper or whiteboards were used for tasks such as implementing new features, fixing bugs, and reviewing design, and drawing tools were also used.
Yatani~et~al.~\cite{yatani2009CHI} conducted questionnaire interviews with nine Ubuntu developers to investigate how developers use diagrams in Ubuntu development.
The interviews revealed that diagrams were used for such as code designing and refactoring, ad-hoc meeting, onboarding, and documentation. The results also showed that it is common case that developers use digital diagrams rather than analog diagrams in the development of Ubuntu.

\noindent
 \textbf{Techniques for Generating Code from Diagrams.}
Studies have been conducted on techniques for generating code from diagrams \cite{supaartagorn2017ICSE,wu2011JSEA,zejie2022EMNLP,mengliang2025ACL}.
Supaartagorn~\cite{supaartagorn2017ICSE} proposed a tool to analyze flowcharts and automatically generate Java and PHP source code.
The tool was evaluated by a group of experts and a group of general users in terms of functional requirement, function, usability, performance and security. As a result, the expert group scored 4.48 out of 5 points, and the general user group scored 4.27.
Wu~et~al.~\cite{wu2011JSEA} proposed a structure identification algorithm that automatically identifies loops and selections according to the structure of flowcharts. They evaluated the correctness and effectiveness of this algorithm using 12 structures, including complex nested structures.

\noindent
 \textbf{Research Questions.}
Although various studies on LLMs for code generation have been conducted, the potential impact of visual design to guide multimodal LLMs for code generation has not been investigated.
In addition, as mentioned in Section 2.3, flowcharts and other diagrams are often created and widely used in software development.
The level of detail varies from simple handwritten flowcharts on a whiteboard to more complex flowcharts. Previous research~\cite{Kaixin2024EMNLP} investigates the ability of multimodal LLMs to generate code for algorithmic challenges that include visual elements. Thus, we propose that employing flowcharts as prompts and providing them to LLMs alongside text could enhance the effectiveness of automatic code generation for coding challenges that contain only textual descriptions.

Therefore, we set the following Research Questions (RQs) to explore the potential of flowcharts on code generation using GPT-4o.

\RqOne

While prior studies have investigated the effectiveness of LLMs in code generation, they mainly focus on textual prompts. In software development, flowcharts are commonly used in design to clarify logic and structure. This question aims to determine whether incorporating such diagrams into the prompt can enhance the LLM's understanding of the problem, thereby improving the performance of code generation. If flowcharts are beneficial, they could be an effective, low-cost addition to LLM-powered development workflows.

\RqTwo

In software engineering practice, flowcharts vary significantly in granularity, from rough high-level sketches to precise fine-grained diagrams. This variability raises a crucial question for multimodal LLMs: is a highly detailed flowchart necessary for effective performance, or is a simpler, more abstract representation sufficient? Answering this question will help developers and researchers in balancing the manual effort required to create detailed diagrams against the potential benefit of enhanced code generation, thus facilitating human-AI collaboration.

\RqThree

Few-Shot Learning has demonstrated significant improvements in text-based tasks by providing the model with several examples~\cite{mingyang2024ICSE,vanhoang2023ICSE,noor2023ICSE}. This question investigates whether providing examples that include both flowcharts and corresponding code can further improve LLM performance in generating correct code from unseen flowcharts. The answer will reveal whether multimodal Few-Shot Learning is a viable strategy for enhancing diagram-based code generation.

\RqFour

As the size of LLMs has grown exponentially, the cost of applying these models has also increased significantly. While larger models often yield superior results, their practical deployment hinges on a critical trade-off between performance and cost. This RQ aims to show this trade-off by analyzing how different prompting strategies affect both performance and cost, enabling users to make more informed, cost-effective decisions.

\section{Experimental Setting}
\label{sec:method}

Figure~\ref{method_flow} shows an overview of the experimental process. This experiment adheres to the methodology outlined by Koyanagi~et~al.~\cite{koyanagi2024MSR}, specifically regarding the construction of datasets, code generation, the execution of tests, and the evaluation of results. The experiments were conducted through the following steps.

\textbf{LLM Selection.}
We select \texttt{GPT-4o} (gpt-4o-2024-08-06) in this experiment. This decision is based on its extensive usage as a prominent language model in the field of code generation research~\cite{dominik2025Access,bruni2025FORGE}. Moreover, it is a state-of-the-art multimodal LLM, which allows it to process both texts and images.

\begin{table}[t]
	\centering
	\caption{Distributions of collected coding questions under different depth settings.}
	\label{tab:num_answer}
	\begin{tabular}{lrrrrr}
		\toprule
		    & A  & B  & C  & D  & Sum  \\
		\midrule
		depth 0 & 92 & 91 & 59 & 47 & 289 \\
		depth 1 & 46 & 71 & 58 & 45 & 220 \\
		depth 2 & 10 & 44 & 38 & 40 & 132 \\
		\bottomrule
	\end{tabular}
\end{table}

\textbf{Dataset Construction.}
Following Koyanagi~et~al.~\cite{koyanagi2024MSR}, we use the coding questions from AtCoder, the largest competitive programming platform in Japan. Our selection process for these questions involves:
(i) inclusion in the AtCoder Beginner Contest (ABC),\footnote{\url{https://atcoder.jp/contests/abc193/tasks/abc193_c?lang=en}} which targets novice programmers, featuring difficulty levels ranging from A to D (with A being the easiest and D the most challenging); (ii) availability of solution code from the official explanation page and test cases from Koyanagi~et~al.~\cite{koyanagi2024MSR}; (iii) the solution code is written in Python.

In summary, we collected a total of 289 coding questions along with their solution code and corresponding test cases for validation. The distribution of difficulty levels among the collected questions is detailed in Table~\ref{tab:num_answer}. The most common coding questions are at difficulty levels A and B, 92 and 91 instances, respectively.

\begin{figure}[t]
\centering
\begin{minipage}{0.48\linewidth}
    \begin{lstlisting}[caption=Example of solution code, label=code:example_answer_code, language=Python]
N = int(input())           # depth 0
sq = int(N ** 0.5)         # depth 0
s = set()                  # depth 0
for a in range(2, sq + 1): # depth 1
    x = a * a              # depth 1
    while x <= N:          # depth 2
    	s.add(x)           # depth 2
    	x *= a             # depth 2
print(N - len(s))          # depth 0
	\end{lstlisting}
\end{minipage}
	\hfill
	\begin{minipage}{0.48\linewidth}
    \includegraphics[width=\linewidth]{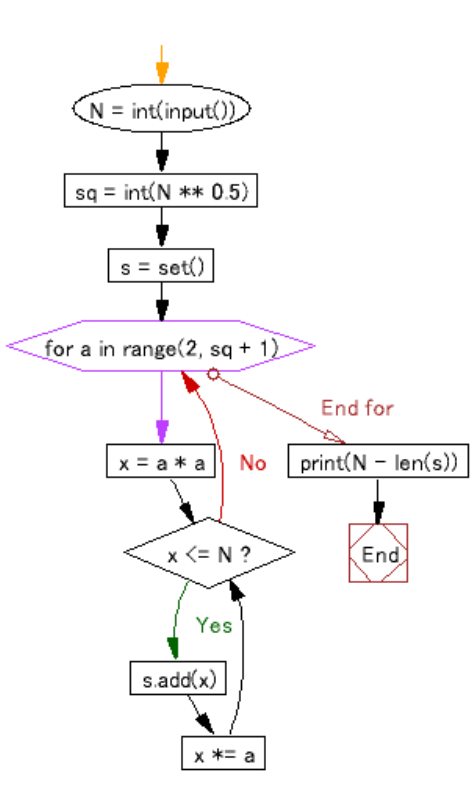}
	\caption{Flowchart generated from Source Code \ref{code:example_answer_code}}
	\label{fig:example_flowchart}
    \end{minipage}
\end{figure}

\textbf{Creation of Flowcharts from Solution Code.}
In our study, we explore the \textit{feasibility or potential} of incorporating flowcharts to enhance code generation. It is important to note that AtCoder does not offer flowcharts to aid programmers in comprehending the given coding questions. To address this, we employ Visustin,\footnote{\url{https://www.aivosto.com/visustin.html}} an automatic tool that generates flowcharts from provided code by leveraging its abstraction details and has been used in the code generation task~\cite{mengliang2025ACL}. Figure~\ref{fig:example_flowchart} illustrates a flowchart produced by Visustin from the Source Code \ref{code:example_answer_code}. The rationale for creating the flowchart from the solution code lies in mimicking developers' practice in coding with high-level design diagrams, thereby we can assess LLMs on code generation tasks for providing flowcharts at varying levels of detail (i.e., from structure design to complete flowchart).

\textbf{Code Generation.} In the following, we present our approach to code generation.

\noindent \textit{Prompt Construction.}
To mimic developers in software development, we designated the role of programming contestant to the LLM within the system prompt as follows.
\begin{itembox}[l]{System prompt}
	You are an excellent competitive programming contestant.
\end{itembox}

To examine the potential impact of providing flowcharts on code generation performance, we developed the following user prompt, accompanied by the problem statement. The content of user prompts differs for each RQ according to its intended purpose.
\begin{itembox}[l]{User prompt}
	The problem statement of a programming task will be given. Output the solution code. The solution code should read from standard input and display the answer as output. Only output the code, without any explanation.

	[Problem Statement of each question]

	[Flowchart created by the solution code for each question]
\end{itembox}

In AtCoder, a problem statement concisely describes the task that needs to be solved using code.  
An example of the problem statement for Source code~\ref{code:example_answer_code} is shown below.

\begin{breakableitembox}[l]{Example of problem statement}
{\footnotesize
Given is an integer N. How many integers between 1 and N (inclusive) are unrepresentable as $a^b$, where a and b are integers not less than 2?\\
\rule{\textwidth}{0.0001pt}
Constraints\\
N is an integer. $1 \le N \le 10^{10}$\\
\rule{\textwidth}{0.0001pt}
Input\\
Input is given from Standard Input in the following format: N\\
\rule{\textwidth}{0.0001pt}
Output\\
Print the answer.\\
\rule{\textwidth}{0.0001pt}
Sample Input 1\\
8\\
\rule{\textwidth}{0.0001pt}
Sample Output 1\\
6\\
4 and 8 are representable as $a^b$: we have $2^2 = 4$ and $2^3 = 8$.\\
On the other hand, 1, 2, 3, 5, 6, and 7 are unrepresentable as $a^b$ using integers a and b not less than 2, so the answer is 6.
}
\end{breakableitembox}

\noindent \textit{Inherent Randomness.}
Given that the code produced by LLMs varies from one instance to another due to inherent randomness \cite{shuyin2025TOSEM}, we executed the code generation process a total of five times to account for this variability.

\noindent \textit{Handling Cases Where Code Generation Cannot Be Performed Correctly.}
In rare cases, code generation may fail due to random problems. In those, code generation will be performed up to three times to ensure the code is correctly generated.

\textbf{Evaluation of Code}
In this step, we run test cases to validate the generated code. During testing, the input examples from the test cases were provided as standard input, and the code was executed to generate the output for each test case scenario. The test execution employed Online-Judge-Tools,\footnote{\url{https://github.com/online-judge-tools/oj}} a Python library used to assess both the correctness of test cases and compliance with time and memory constraints.

In this experiment, we employ the \textit{Accuracy} evaluation metric, represented as the percentage of correct solutions. \(\textit{Accuracy}_i\) is determined for each \(i\)th generation from varying difficulty levels. Specifically, \(\textit{Accuracy}_i\) quantifies the proportion of code solutions that successfully passed all test cases as input among the total solutions generated in the \(i\)th generation from different difficulties. Given that RQ2 to RQ3 require setting different depths for a solution code, resulting in a varying number of coding questions for each prompt configuration, we ensure fair comparisons by reporting the accuracy from 132 questions specifically at a depth of 2 in Table~\ref{tab:num_answer}.

\section{Flowcharts for Code Generation}
\label{sec:RQ1}
Since software diagrams can be used as comprehension aids~\cite{cross1998EMSE,hungerford2004TSE} to supplement textual descriptions in software development, they can also help LLMs understand the requirements for code generation. In this section, we validate whether providing flowcharts can improve the accuracy of code generation.

\subsection{Approach}
To answer RQ1, we set two prompting strategies to examine how including FlowCharts (FC) together with the Problem Statement (PS) affects accuracy: (i) PS only, and (ii) PS combined with FC.

By providing only the textual description of coding questions, we set LLMs to mimic developers in writing code without drawing any design diagrams. For another prompting strategy, we incorporate Visustin-generated flowcharts to guide coding for these questions. Therefore, we provide the following instruction to detail the flowchart within the user prompt.
\begin{itembox}[l]{Description of the flowchart}
	The attached image is an example of the solution codes for this problem.
\end{itembox}

\begin{table}[t]
	\centering
	\caption{Median \textit{Accuracy} for each difficulty level and average \textit{Accuracy} for each prompting strategy across all difficulty levels.}
	\label{tab:median_accuracy_table}
	\scalebox{0.80}{
		\begin{tabular}{lcrrrrr}
			\toprule
			& Prompting Strategy & A & B & C & D & Average \\
			\midrule
			\multirow{2}{*}{\rotatebox[origin=c]{90}{\textbf{RQ1}}} & PS 	& 90.00 & 90.91 & 76.32 & 42.50 & 74.93 \\
			\cmidrule{2-7}
			& PS + FC & \textbf{100.00} & \textbf{95.45} & \textbf{89.47} & \textbf{55.50} & \textbf{85.11} \\
			\midrule
			\multirow{4}{*}{\rotatebox[origin=c]{90}{\textbf{RQ2}}} &  0-Shot PS + FC (abstracted at depth 0) & 80.00 & 88.64 & 81.58 & 37.50 & 71.93 \\
			\cmidrule{2-7}
			& 0-Shot PS + FC (abstracted at depth 1) & 80.00 & 88.64 & 84.21 & 47.50 & 75.09 \\
			\cmidrule{2-7}
			& 0-Shot PS + FC(abstracted at depth 2) & 100.00 & 93.18 & 86.84 & 52.50 & 83.13 \\
			\midrule
			\multirow{8}{*}{\rotatebox[origin=c]{90}{\textbf{RQ3}}}
            & 0-Shot PS + FC(abstracted at depth 1) & 80.00 & 88.64 & 84.21 & 47.50 & 75.09 \\
			\cmidrule{2-7}
            & 1-Shot PS + FC(abstracted at depth 1)  & 90.00 & 93.18 & 89.47 & 55.00 & 81.91 \\
			\cmidrule{2-7}
            & 2-Shot PS + FC(abstracted at depth 1)  & 100.00 & 88.64 & 86.84 & 42.50 & 79.50 \\
			\cmidrule{2-7}
            &  0-Shot PS + FC(abstracted at depth 2)  & 100.00 & 93.18 & 86.84 & 52.50 & 83.13 \\
			\cmidrule{2-7}
			& 1-Shot PS + FC(abstracted at depth 2)  & 100.00 & 93.18 & 89.47 & 52.50 & 83.79 \\
			\cmidrule{2-7}
			& 2-Shot PS + FC(abstracted at depth 2) & 100.00 & 93.18 & 89.47 & 55.00 & 84.41 \\
			\bottomrule
		\end{tabular}
	}
\end{table}

\begin{figure*}[t]
	\centering
	\includegraphics[width=\linewidth]{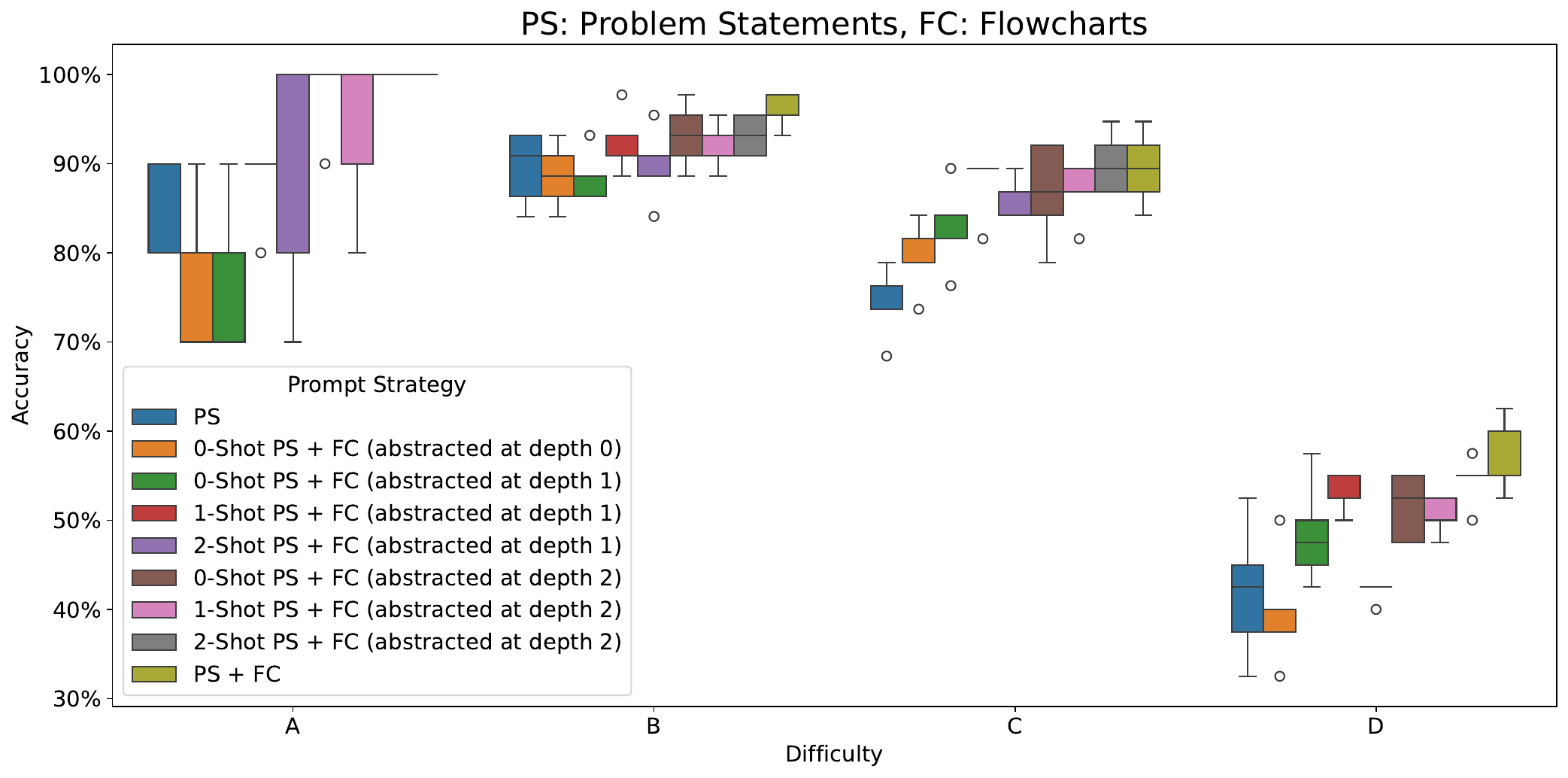}
	\caption{Accuracy of different prompting strategies}
	\label{fig:accuracy_boxplot}
\end{figure*}

\subsection{Results}
Figure~\ref{fig:accuracy_boxplot} shows the accuracy variations with various prompting strategies for different coding difficulties. Table~\ref{tab:median_accuracy_table} presents the distinct accuracy variations using various prompting strategies across coding questions, highlighting differences in median and average.

\textbf{Observation 1. Providing flowcharts can substantially improve LLM accuracy in code generation tasks.}
By providing only textual descriptions of coding questions, the LLM achieves an average accuracy of 74.93\% on AtCoder coding questions, indicating the LLM's promising performance in code generation.
The PS + FC prompting strategy yields the highest accuracy in code generation at all difficulty levels, accounting for the average accuracy of 85.11\% in the top portion of Table~\ref{tab:median_accuracy_table}. The difference between the two prompting strategies indicates that the flowchart not only assists developers in comprehension but also guides LLMs in code generation.

\textbf{Observation 2. The performance gap from incorporating flowcharts in prompts increases as the difficulty level increases.}
In Table~\ref{tab:median_accuracy_table}, we observed a 4--10\% increase in code generation for simpler coding questions A and B, while PS + FC achieved a 13\% increase for more challenging questions like C and D. This suggests that high-level design diagrams, such as flowcharts, enhance LLM comprehension, especially in complex coding questions.

\conclusionbox{
	\textbf{RQ1 Summary:}
	Providing high-level diagrams, such as flowcharts, not only facilitates a thorough comprehension for developers but also enhances multimodal LLM accuracy in code generation. This improvement can be notably significant for more complex coding questions, ranging from 13\%.
    }

\section{Abstracted Flowcharts for Code Generation}
\label{sec:RQ2}
In RQ1, our experiments provide evidence that incorporating flowcharts as visual designs to supplement textual requirements can significantly improve the accuracy of code generation. In this RQ, we investigate the trade-off between the details in the flowcharts (manual efforts required to create them) and the accuracy of code generation.

\subsection{Approach}
In Python-written solution code, control structures such as \textit{if}, \textit{for}, and \textit{while} often contain the core programming logic needed to solve the problem.
Therefore, we posit that it is feasible to create flowcharts with different levels of abstraction using the indentation associated with internal control structures.

The initial step in producing the flowchart with abstraction levels involves establishing the depth of each line within the solution code.
The abstraction of the flowchart is achieved by replacing the sections of the solution code that exceed a specified depth. To build the abstraction of the flowchart, we set a depth of 0 for global variable definitions and import statements in the given solution code.
For control structures such as function and class definitions, \textit{if}, \textit{for}, and \textit{while}, we set the depth one level deeper to the definition expressions,  conditional expressions, iterator expressions and statements within the indented body.
Then, as new control structures appear, the depth of control structures is added in increments of 1.
Source Code \ref{code:example_answer_code} shows an example of the depth setting with the level of depth annotated by comments within the Python code.

The abstraction is applied to segments of the solution code that exceed this predetermined depth. This process modifies the detail level of the flowcharts.
The extraction is performed based on the following strategies, where Algorithm \ref{code:abstraction_algorithm} illustrates the pseudocode of the abstraction algorithm.

\begin{algorithm}[tb]
    \caption{Abstraction Solution Codes}
    \scriptsize
    \label{code:abstraction_algorithm}
    \begin{algorithmic}[1]
\Function {abstract\_code}{node, abstraction\_depth}
    \If{abstraction\_depth $\le$ node.depth}
        \If{node.type $\in$ \{FunctionDef, ClassDef, If, For, While\}}
            \State node.body $\gets$ Pass
        \EndIf
        \If{node.type $\in$ \{FunctionDef, ClassDef\}}
            \State node.args $\gets$ [ ]
        \EndIf
        \If{node.type $\in$ \{If, While\}}
            \State node.condition $\gets$ None
        \EndIf
        \If{node.type = For}
            \State node.iter $\gets$ None
        \EndIf
    \EndIf
    \If {abstraction\_depth == 0}
        \If{node.type = Assign}
            \If{\textbf{not} is\_input\_expr(node.value)}
                \If{type(node.value) $\in$ \{int, str\}}
                    \State node.value $\gets$ None
                \ElsIf{type(node.value) = list}
                            \State node.value $\gets$ [ ]
                \ElsIf{type(node.value) = tuple}
                    \State node.value $\gets$ ( )
                \EndIf
            \EndIf
        \EndIf
        \If{node.type = FunctionCall}
            \State node.args $\gets$ [ ]
        \EndIf
    \EndIf
    \For{child $\in$ node.children}
        \State \Call{abstract\_code}{child, abstraction\_depth}
    \EndFor
\EndFunction

\Function{is\_input\_expr}{expr}
    \State \Return contains\_input(expr) \Comment{e.g., input(), sys.stdin, etc.}
\EndFunction
    \end{algorithmic}
\end{algorithm}

\begin{itemize}
    \item If the depth of an \textit{if}, \textit{for}, \textit{while}, function, or class definition exceed the abstraction depth, the indented body is replaced with \textit{pass}. Also, in the case of function and class definitions, the arguments in the definition expressions are removed, and within \textit{if} and \textit{while} statements, conditional expressions, as well as the iterator expressions in \textit{for} statements, are replaced by \textit{None}.
    \item Regarding the abstraction of rows at a depth of 0, we adopted a distinct approach depending on the following scenarios in addition to the abstraction at depth 1.
    \begin{itemize}
        \item For assignment statements, variable definitions with the type \textit{int} or
        \textit{str} are replaced by \textit{None} on the right-hand side. Variable definitions with the type \textit{list} or \textit{tuple} are replaced by an empty \textit{list} or \textit{tuple}.
        \item If the assignment statement involves processing of inputs through an input function, etc., that part remains unabstracted.
        \item When a function call is present, the argument is removed.
    \end{itemize}
\end{itemize}


\begin{figure}[t]
\centering
\begin{minipage}{0.48\linewidth}
    \begin{lstlisting}[caption=Abstraction of Source Code A at depth 0, label=code:example_abstracted_code, language=Python]
N = int(input())
sq = int()
s = set()
for a in None:
	pass
print()
	\end{lstlisting}
\end{minipage}
	\hfill
	\begin{minipage}{0.48\linewidth}
    \centering
    \includegraphics[width=.3\linewidth]{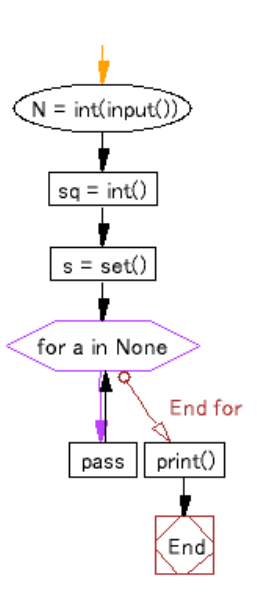}
	\caption{Abstracted flowchart at depth 0 from Source Code \ref{code:example_abstracted_code}}
	\label{fig:example_abstracted_flowchart}
    \end{minipage}
\end{figure}

Source Code \ref{code:example_abstracted_code} presents the abstracted code at depth 0 for Source Code \ref{code:example_answer_code}.
Figure \ref{fig:example_abstracted_flowchart} shows the abstracted flowchart generated from Source Code \ref{code:example_abstracted_code}.

We abstracted the solution code from depths ranging from 0 to 2. The maximum depth of a solution code differs depending on the code, and code for simpler problems typically has a shallower depth. Consequently, increasing the depth reduces the number of solution code that can be abstracted. Table \ref{tab:num_answer} shows the number of coding questions available in different depth settings. In total, we have 132 coding questions for a depth of 2 in abstraction.

In \textbf{RQ2}, the user prompts
are formulated under three prompting strategies to examine how various depth settings of flowcharts affect accuracy: (iii) PS + FC (abstraction at depth 0); (iv) PS + FC (abstraction at depth 1); and (v) PS + FC (abstraction at depth 2).

We then include the following instruction in the task description section of the user prompt.
\begin{itembox}[l]{Description of the abstracted flowchart}
	The attached image is an abstract flowchart of the solution code idea for this problem.
\end{itembox}

\subsection{Results}
\textbf{Observation 3. Providing more information in flowcharts enhances the accuracy of the LLM in code generation.}
The middle portion of Table~\ref{tab:median_accuracy_table} shows the accuracy variations under different depth settings of flowcharts. Overall, the LLM surpasses accuracy by merely offering problem statements ($74.93 > 71.93$) when the solution code contains only global variable definitions and import declarations (depth of 0). In contrast, incorporating depth of 2 resulted in an increase of approximately 8\%.

\textbf{Observation 4. Providing complete flowcharts results in only a marginal enhancement of code generation compared to the abstraction at a depth of 2.}
Our analysis reveals that the model, when configured to a depth of 2, outperforms other configurations with an average accuracy of 83.13\%. Although this value does not exceed the results obtained from using the complete flowchart (PS + FC), the performance gain is marginal, around 2\%. Providing flowcharts with a depth of 1 incorporating the most outlier control structures only enhances marginal performance; accuracy increases by 0--3\% for questions at levels A--C. Conversely, for the most complex questions, this configuration yields an improvement of 10\%.

\conclusionbox{
	\textbf{RQ2 Summary:}
    Providing more information in flowcharts generally enhances the accuracy of LLMs in code generation. Models configured to a depth of 2 achieve a significant average accuracy of 83.13\%. However, while including global variable definitions and import statements marginally improves accuracy, it varies with question complexity, improving complex question accuracy but not simpler ones.
}

\section{Few-Shots for Code Generation}
\label{sec:RQ3}
In this section, we explore the impact of Few-Shot Learning on code generation when using abstracted flowcharts.

\subsection{Approach}
When incorporating flowcharts in the prompt, we also apply 1-Shot, which provides a single input/output example, and 2-Shot, which provides two input/output examples.
The abstraction strategies chosen for RQ3 demonstrated an improvement in \textit{Accuracy} compared to the use of problem statements. The configuration of the user prompt is structured as follows:

\begin{itembox}[l]{User prompt}
	The problem statement of a programming task will be given. Output the solution code. The solution code should read from standard input and display the answer as output. Only output the code, without any explanation.

	[Input Example]

	[Output Example]

	[Problem Statement of each question]

	[Flowchart created by the solution code for each question]
\end{itembox}

In the ``Input Example'' section, we incorporated the problem statement alongside a flowchart. In the ``Output Example'' section, we provide the solution code corresponding to the added questions. The level of abstraction for the flowchart included as the ``Input Example'' should be consistent in \textbf{RQ2}. As a reference, we included questions resembling the targeted problem statement for code generation. We used 289 coding questions as candidates to calculate cosine similarity. Initially, the cosine similarity between the problem statements for code generation and those of other questions is computed using \texttt{Sentence Transformers}.\footnote{\url{https://sbert.net/}} The coding questions exhibiting the highest cosine similarity are used for 1-Shot, whereas the coding questions with the highest and second-highest cosine similarities are employed for 2-Shot.

In RQ2, we found that abstracted flowcharts at depths of 1 to 2 outperform those that provide only a problem statement in complex tasks. Therefore, in RQ3, we investigated the effectiveness of Few-Shot Learning when applied to prompts using flowcharts abstracted at depths 1 and 2. We set four different prompting strategies:
(vi) 1-Shot PS + FC (abstraction at depth 1); (vii) 2-Shot PS + FC (abstraction at depth 1); (viii) 1-Shot PS + FC (abstraction at depth 2); and (ix) 2-Shot PS + FC (abstraction at depth 2).

\subsection{Results}
\textbf{Observation 5. One-Shot Learning has a substantial improvement in code generation compared to those abstract flowcharts at depth 1 with Zero-Shot Learning.} As illustrated in the bottom portion of Table~\ref{tab:median_accuracy_table} and Figure~\ref{fig:accuracy_boxplot}, our analysis reveals that One-Shot Learning causes an improvement (6\%) on average at a depth of 1. An abstract flowchart at a depth of 1, along with the most similar example, can enhance code generation by 4--10\%. Among nine prompting strategies, it balances the effort of creating a flowchart (abstracting at a depth of 1) with the accuracy of the generated code.

\textbf{Observation 6. Few-Shot Learning that has more than one example only produces a marginal enhancement of code generation.}
Conversely, increasing the number of examples resulted in a decrease in accuracy from 81.91\% to 79.50\%. At depth 2, this trend did not manifest; instead, accuracy slightly improved from 83.13\% in Zero-Shot to 84.41\% in Two-Shot. For questions at levels A--B, the accuracy remains unchanged.

\conclusionbox{
    \textbf{RQ3 Summary:}
    Few-shot learning, particularly one-shot, demonstrated improved accuracy for solving coding challenges using flowcharts. However, offering additional examples beyond one can decrease accuracy or lead to marginal improvement.
}

\section{Cost of Flowcharts for Code Generation}
\label{sec:RQ4}
In this section, we investigate how providing flowcharts to prompts affects the number of tokens and costs when using the \texttt{GPT-4o} API.

\subsection{Approach}

The number of input tokens was retrieved from the OpenAI API, which provides the number of tokens used in the input prompt; the number of output tokens was calculated using \texttt{tiktoken},\footnote{\url{https://github.com/openai/tiktoken}} a library provided by OpenAI. We referred to the price list for the \texttt{GPT-4o} batch API\footnote{\url{https://platform.openai.com/docs/pricing?latest-pricing=batch}} to calculate the cost of input and output tokens. Specifically, the cost of input tokens is \$0.00000125 per token, and the cost of output tokens is \$0.000005 per token.
If code generation is performed again for questions that could not be generated, the tokens and costs used in that generation will be used as results.

\subsection{Results}
\label{subsec:token_cost}
\begin{table*}[t]
    \centering
    \caption{Number of tokens and cost used in all experiments (left) and per code generation (right).}
    \label{tab:token_and_cost}
    \resizebox{.95\textwidth}{!}{
    \begin{tabular}{crrrrrrr!{\vrule width 1pt\hspace{1pt}\vrule width 1pt}rrrrrr}
        \toprule
        & & \multicolumn{3}{c}{Token (Total)} & \multicolumn{3}{c}{Cost (\$, Total)} &
        \multicolumn{3}{c}{Token (Per Generation)} & \multicolumn{3}{c}{Cost (\$, Per Generation)} \\
        \cmidrule(lr){3-5} \cmidrule(lr){6-8} \cmidrule(lr){9-11} \cmidrule(lr){12-14}
        \makecell{Prompting \\Strategy} & \makecell{Number of \\Generations} &
        Input & Output & Total & Input & Output & Total &
        Input & Output & Total & Input & Output & Total \\
        \midrule
        PS & 1,445 & 580,375 & 199,615 & 779,990 & 0.73 & 1.00 & 1.73 & 401.64 & 138.14 & 539.78 & 0.00050 & 0.00069 & 0.00119 \\
        \midrule
        PS + FC & 1,445 & 1,191,515 & 131,794 & 1,323,309 & 1.49 & 0.66 & 2.15 & 824.58 & 91.20 & 915.78 & 0.0010 & 0.00045 & 0.00145 \\
        \midrule
        \makecell{PS + FC\\(abstracted at depth 0)} & 1,445 & 1,036,458 & 140,305 & 1,177,673 & 1.30 & 0.70 & 2.00 &
        717.27 & 97.10 & 814.37 & 0.00090 & 0.00049 & 0.00139 \\
        \midrule
        \makecell{0-Shot PS + FC\\(abstracted at depth 1)} & 1,100 & 838,649 & 125,015 & 964,540 & 1.05 & 0.63 & 1.68 &
        762.41 & 113.65 & 876.06 & 0.00095 & 0.00057 & 0.00152 \\
        \midrule
        \makecell{1-Shot PS + FC\\(abstracted at depth 1)} & 1,100 & 1,733,690 & 111,599 & 1,845,289 & 2.17 & 0.56 & 2.73 &
        1,576.08 & 101.45 & 1,677.53 & 0.0020 & 0.00051 & 0.00251 \\
        \midrule
        \makecell{2-Shot PS + FC\\(abstracted at depth 1)} & 1,100 & 2,632,190 & 110,026 & 2,742,216 & 3.29 & 0.55 & 3.84 &
        2392.90 & 100.02 & 2,492.92 & 0.0030 & 0.00050 & 0.0035 \\
        \midrule
        \makecell{0-Shot PS + FC\\(abstracted at depth 2)} & 660 & 619,563 & 90,979 & 710,542 & 0.77 & 0.45 & 1.22 &
        938.73 & 137.85 & 1,076.58 & 0.0012 & 0.00069 & 0.00189 \\
        \midrule
        \makecell{1-Shot PS + FC\\(abstracted at depth 2)} & 660 & 1,293,745 & 86,953 & 1,380,698 & 1.62 & 0.43 & 2.05 &
        1,960.21 & 131.75 & 2,091.96 & 0.0025 & 0.00066 & 0.00316 \\
        \midrule
        \makecell{2-Shot PS + FC\\(abstracted at depth 2)} & 660 & 1,994,450 & 88,104 & 2,082,554 & 2.49 & 0.44 & 2.93 &
        3,021.89 & 133.49 & 3,155.38 & 0.0038 & 0.00067 & 0.00447 \\
        \bottomrule
    \end{tabular}
    }
\end{table*}

Table \ref{tab:token_and_cost} shows the number of input tokens and output tokens when generating code using GPT-4o with each prompting strategy, with the cost corresponding to the tokens used.

\textbf{Observation 7. The input cost increases with the addition of flowcharts, but the total cost remains low.}
As shown in Table \ref{tab:token_and_cost}, adding abstracted flowcharts or complete flowcharts at depths 0, 1, and 2 to the prompt increased the number of input tokens and input costs by approximately 1.45 to 2.34 times.
When performing Few-Shot Learning, the number of input tokens and input costs increase linearly compared to Zero-Shot Learning. In the PS + FC (abstracted at depth 1) strategy, the number of input tokens and input costs increased approximately 2.07 times for 1-Shot and 3.14 times for 2-Shot compared to 0-Shot. In the PS + FC (abstracted at depth 2) strategy, the number of output tokens and output costs increased approximately 2.09 times in 1-Shot and 3.22 times in 2-Shot compared to 0-Shot.
However, the total cost (input cost + output cost) remains low, around \$1.22 to \$3.84 per code generation, even when using flowcharts. This is because the input cost is relatively low compared to the output cost, and the increase in input tokens does not significantly affect the total cost.

\textbf{Observation 8. The output costs did not increase even when the flowchart was added, but rather tended to decrease.}
Table \ref{tab:token_and_cost} shows that, in strategies with the flowchart added, the number of output tokens and output costs was generally lower than in the PS strategy.
In particular, the PS + FC (abstracted at depth 0) strategy reduced the number of output tokens and output cost to approximately 70.29\% of the PS strategy, and the PS + FC strategy reduced them to approximately 66.02\% of the PS strategy. This is likely because the PS + FC (abstracted at depth 0) strategy generates shorter code due to the limited information in the flowchart, while the PS + FC strategy generates more refined code by directly incorporating the solution code information.
Additionally, when performing Few-Shot Learning, there is a tendency for the number of output tokens and output costs to decrease slightly compared to when performing Zero-Shot Learning using the same flowchart. In the PS + FC (abstracted at depth 1) strategy, the 1-Shot method achieved 89.27\% of the 0-Shot method, and the 2-Shot method achieved 88.01\% of the 0-Shot method. In the PS + FC (abstracted at depth 2) strategy, 1-Shot was 95.57\% of 0-Shot, and 2-Shot was 96.84\% of 0-Shot.

\conclusionbox{
    \textbf{RQ4 Summary:}
    Using flowcharts as input in a multimodal LLM for code generation is economically efficient. Although this approach requires the use of more tokens in the input, it results in generating code that is both more concise and precise.
}

\section{Discussion}
\label{sec:discussion}
We discuss the implications derived based on the results and threats to the validity of our study.

\subsection{Implications}

\implication{Using flowcharts to supplement textual requirements is more efficient in terms of both cost and performance, as it closely represents the structure of the code.}
For code generation, the most promising prompting strategy is to provide the problem statement and flowchart. The number of tokens and cost when using this prompting strategy is approximately 1.78 times that of the PS strategy, but in reality, the cost has not increased significantly since it produces more concise and precise code than providing only the problem statement.
Considering the above, it is concluded that using the PS + FC strategy (or a flowchart with a level of detail closer to a complete flowchart) is more efficient in terms of both cost and performance.

\implication{When using LLMs by incorporating flowcharts in practical software development, flowcharts with one level of depth and one example close to the problem should be used.}
Based on the findings of RQ1--3, the PS + FC strategy exhibited the highest \textit{Accuracy}. However, in real-world development contexts, if a flowchart depicting the complete code structure is available, directly writing the complete code could potentially be faster. This is because the capability to generate a complete flowchart indicates that the developer already has the corresponding code structure mentally. In addition, flowcharts generated at a depth of 0 did not considerably enhance \textit{Accuracy}, whereas producing those abstracted flowcharts at depths 1 and 2, or complete flowcharts, led to better \textit{Accuracy}.

Furthermore, our Few-Shot Learning experiments suggest that providing an example similar to coding questions results in more accurate code generation. Consequently, it is advisable for developers or researchers to provide an example and create flowcharts with a depth of 1, detailing the overall structure, including control structures like variable definitions and if statements, and to automatically employ one example similar to the problems. This illustrates the trade-off between manual effort to produce diagrams and the guidance offered to LLMs for code generation.

\implication{Future work is required to evaluate the quality of generated code beyond correctness.} Building upon the insights from RQ3, which showed that generated flowcharts guide LLMs in producing more precise and concise code (resulting in shorter code), our future work will extend this observation beyond functional correctness. We aim to systematically evaluate the readability, maintainability, and robustness of code generated through various prompting strategies from abstracted flowcharts, towards employing design diagrams to produce high-quality code.

\subsection{Threats to Validity}
\label{sec:threats_to_validity}

\noindent \textbf{Construct Validity.}
The RQ1 results show that the PS + FC prompting strategy yields the highest \textit{Accuracy}, and that \textit{Accuracy} increases as flowcharts become less abstract and more detailed. However, using abstract flowcharts in real development may be impractical, since time spent on them could instead be used to write the target code.
 If a flowchart closely matches the final code structure, it may be more efficient to write the code directly rather than use the flowchart as a prompt. 
In addition, our controlled experiments may not reflect real-world software development workflows, where developers typically create flowcharts before coding, so the observed effectiveness should be interpreted cautiously when generalizing to practical settings. 

In this paper, we evaluated generated code solely in terms of \textit{Accuracy}. While functional correctness is a widely used metric, it does not capture other important aspects of code quality, such as readability, maintainability, and coding style. As a result, our evaluation may not fully represent the overall quality of the generated code.
Furthermore, while generating flowcharts from solution code introduces a risk of information leakage, this design was essential for our feasibility study due to the lack of available flowchart datasets for the studied coding questions. Our primary goal was to explore whether structural representations like flowcharts can clarify problem logic for LLMs. To address this limitation, future work will focus on creating flowcharts directly from natural language problem descriptions without referencing the solution code by employing developers.

\noindent \textbf{Internal Validity.}
In RQ2, we generated code using abstracted flowcharts and evaluated it. When abstracting, the indentation width, or depth value, differs for each solution code, so the number of problems used between prompts varies. Therefore, it may be difficult to strictly compare \textit{Accuracy} between each prompting strategy.
Additionally, we discussed token count and cost, arguing that using a flowchart could potentially reduce output token count and output cost. However, in this experiment, we adjusted the output to generate only code. If the output includes text other than code, using a flowchart could potentially increase the output token count and output cost.

\noindent \textbf{External Validity.}
This experiment was conducted using AtCoder problems; therefore, the results may differ for problems from other competitive programming contests (e.g., LeetCode~\cite{Kaixin2024EMNLP}) or for real-world software tasks (e.g., web apps). In this paper, we evaluated only GPT-4o, one of the most capable multimodal LLMs available at the time of our study; other multimodal LLMs may behave differently and show different levels of improvement with flowcharts. Thus, our findings may not generalize to other tasks or LLMs.

\section{Conclusion}
\label{sec:conclusion}

This study investigates the effect of program flowcharts as visual prompts to enhance the performance of automatic code generation by multimodal Large Language Models (LLMs). To quantify this effect, we conducted a systematic evaluation using coding questions from AtCoder, a popular competitive programming site in Japan. Our methodology involved comparing nine prompting strategies to assess their impact on the accuracy and cost of code generation through the use of the multimodal LLM API (\texttt{GPT-4o}). Specifically, we evaluated the four main aspects: (i) providing only the textual problem statement as a baseline; (ii) incorporating the problem statement with a flowchart generated from the solution code to represent the high-level design in the coding question; (iii) incorporating the problem statement with a flowchart generated from an abstracted version of the solution code; and (iv) applying Few-Shot Learning by providing examples related to the coding questions to each of the aforementioned aspects.

Our results reveal significant improvements in the accuracy of the generated code by incorporating flowcharts as the high-level design. This improvement was particularly pronounced for more challenging problems, where accuracy increased approximately 10\%. The most effective prompting strategy was the combination of the problem statement with a flowchart generated from the original solution code, which consistently yielded the highest accuracy. Flowcharts that were highly abstracted provided minimal or no performance benefit, suggesting that the level of detail in the visual prompt is a critical factor for performance. Moreover, the application of Few-Shot Learning, especially when combined with abstracted flowcharts, results in performance gains for one-shot learning but only marginal gains for two-shot learning. Furthermore, incorporating flowcharts in code generation is cost-effective.

Our findings underscore the potential of visual design to guide multimodal LLMs for complex programming tasks, such as code generation. Future research will evaluate enhancement possibilities by incorporating explicit problem difficulty levels into prompts and fine-tuning the model for these tasks. This opens up new research and several promising avenues for exploration, such as experimenting with other forms of high-level diagrams (e.g., UML diagrams) and multimodal inputs for diagram-driven development. A key objective will be to assess the balance between flowchart detail and LLM efficiency across various programming challenges and domains.

\subsubsection{DATA AVAILABILITY.}
The complete replication package can be accessed at \url{https://doi.org/10.5281/zenodo.17360470}.

\begin{credits}
\subsubsection{\ackname}
We gratefully acknowledge the financial support of: (1) JSPS for the KAKENHI grants (JP24K02921, JP25K03100, JP25K22845, JP26H02500, JP26K21198); (2) Japan Science and Technology Agency (JST) as part of Adopting Sustainable Partnerships for Innovative Research Ecosystem (ASPIRE), Grant Number JPMJAP2415, and (3) the Kayamori Foundation of Informational Science Advancement for supporting Tao Xiao, and (4) the Inamori Research Institute for Science for supporting Yasutaka Kamei via the InaRIS Fellowship.

\subsubsection{\discintname}
All co-authors have seen and agree with the contents of the manuscript, and there is no financial interest to report.
\end{credits}

\bibliographystyle{splncs04}
\bibliography{references}

\end{document}